\documentclass[twocolumn,aps,showpacs,floatfix,prc]{revtex4}
\usepackage[dvips]{epsfig}
\begin{document}
\title{ Core-crust transition and crustal fraction of moment of inertia in neutron stars }

\author{ Debasis Atta$^{1*\dagger}$, Somnath Mukhopadhyay$^{2*}$ and D. N. Basu$^{3*}$ }

\affiliation{$^*$ Variable  Energy  Cyclotron  Centre, 1/AF Bidhan Nagar, Kolkata 700 064, India }
\affiliation{ $^\dagger$ Shahid Matangini Hazra Govt. Degree College for Women, Tamluk, West Bengal 721 649, India}

\email[E-mail 1: ]{debasisa906@gmail.com}
\email[E-mail 2: ]{somnathm@vecc.gov.in}
\email[E-mail 3: ]{dnb@vecc.gov.in}

\date{\today }

\begin{abstract}

    The crustal fraction of moment of inertia in neutron stars is calculated using $\beta$-equilibrated nuclear matter obtained from density dependent M3Y effective interaction. The transition density, pressure and proton fraction at the inner edge separating the liquid core from the solid crust of the neutron stars determined from the thermodynamic stability conditions are found to be $\rho_t=$ 0.0938 fm$^{-3}$, P$_t=$ 0.5006 MeV fm$^{-3}$ and $x_{p(t)}=$ 0.0308, respectively. The crustal fraction of the moment of inertia can be extracted from studying pulsar glitches and is most sensitive to the pressure as well as density at the transition from the crust to the core. These results for pressure and density at core-crust transition together with the observed minimum crustal fraction of the total moment of inertia provide a new limit for the radius of the Vela pulsar: $R \geq 4.10 + 3.36 M/M_\odot$ kms.
    
\vskip 0.2cm  
  

\end{abstract}


\maketitle

\noindent
\section{Introduction}
\label{section1}

    Pulsar glitches, which are discontinuities in the spin-down of pulsars, involve sudden transfer of angular momentum from an isolated component (consisting of superfluid neutrons in crust) to the entire star through vortex unpinning. The sudden jumps in rotational frequencies $\omega$ which may be as large as $\frac{\Delta\omega}{\omega} \sim10^{-6} - 10^{-9}$ have been observed for many pulsars. The frequency of observed glitches is statistically consistent with the hypothesis that all radio pulsars experience glitches \cite{Ly96}. Glitches are thought to originate from interactions between the rigid neutron star crust, typically somewhat more than a kilometer thick, and rotational vortices in a neutron superfluid. The inner kilometer of crust consists of a crystal lattice of nuclei immersed in a neutron superfluid \cite{Li99}. Because the pulsar is spinning, the neutron superfluid (both within the inner crust and deeper inside the star) is threaded with a regular array of rotational vortices. As the spin of the pulsar gradually slows, these vortices must gradually move outwards since the rotational frequency of a superfluid is proportional to the density of vortices. Deep within the star, the vortices are free to move outwards. In the crust, however, the vortices are pinned by their interaction with the nuclear lattice. Various theoretical models \cite{Ep92,Al93,Li96,Ru98,Se99} differ in important respects as to how the stress associated with pinned vortices is released in a glitch: for example, the vortices may break and rearrange the crust, or a cluster of vortices may suddenly overcome the pinning force and move macroscopically outward, with the sudden decrease in the angular momentum of the superfluid within the crust resulting in a sudden increase in angular momentum of the rigid crust itself and hence a glitch. All the models agree that the fundamental requirements are the presence of rotational vortices in a superfluid and the presence of a rigid structure which impedes the motion of vortices and which encompasses enough of the volume of the pulsar to contribute significantly to the total moment of inertia.

    In the present work, the equation of state (EoS) used is obtained from the density dependent M3Y effective nucleon-nucleon interaction (DDM3Y) for which the incompressibility $K_\infty$ for the symmetric nuclear matter (SNM), nuclear symmetry energy $E_{sym}(\rho_0)$ at saturation density $\rho_0$, the isospin dependent part $K_\tau$ of the isobaric incompressibility and the slope $L$ are in excellent agreement with the constraints recently extracted from measured isotopic dependence of the giant monopole resonances in even-A Sn isotopes, from the neutron skin thickness of nuclei, and from analyses of experimental data on isospin diffusion and isotopic scaling in intermediate energy heavy-ion collisions \cite{Ch09,Ba09}. The core-crust transition in neutron stars is determined \cite{At14} by analyzing the stability of the $\beta$-equilibrated dense nuclear matter with respect to the thermodynamic stability conditions \cite{La07,Ku04,Ku07,Wo08,Ca85}. The mass-radius relation for neutron stars is obtained by solving the Tolman-Oppenheimer-Volkoff Equation (TOV) \cite{TOV39a,TOV39b} and then the crustal fraction of moment of inertia is determined using pressure and density at core-crust transition. As the angular momentum requirements of glitches in Vela pulsar indicate that $1.4\%$ of the star's moment of inertia drives these events, the allowed region for masses and radii for Vela pulsar is determined from the condition that the crustal fraction of moment of inertia $\frac{\Delta I}{I} > 0.014$ which sets a new limit for its radius.  

\noindent
\section{ Core-crust transition in $\beta$-equilibrated neutron star matter }
\label{section2}

    The nuclear matter EoS is calculated using the isoscalar and the isovector \cite{La62,Sa83} components of M3Y interaction along with density dependence. The density dependence of this DDM3Y effective interaction is completely determined from nuclear matter calculations. The equilibrium density of the nuclear matter is determined by minimizing the energy per baryon. The energy variation of the zero range potential is treated accurately by allowing it to vary freely with the kinetic energy part $\epsilon^{kin}$ of the energy per baryon $\epsilon$ over the entire range of $\epsilon$. This is not only more plausible, but also yields excellent result for the incompressibility $K_\infty$ of the SNM which does not suffer from the superluminosity problem \cite{BCS08}. In a Fermi gas model of interacting neutrons and protons, with isospin asymmetry $X= \frac{\rho_n - \rho_p} {\rho_n + \rho_p},~~~~\rho = \rho_n+\rho_p,$ where $\rho_n$, $\rho_p$ and $\rho$ are the neutron, proton and baryonic number densities respectively, the energy per baryon for isospin asymmetric nuclear matter can be derived as \cite{BCS08}

\begin{equation}
 \epsilon(\rho,X) = \left[\frac{3\hbar^2k_F^2}{10m_b}\right] F(X) + \left(\frac{\rho J_v C}{2}\right) \left(1 - \beta\rho^n\right)  
\label{seqn1}
\end{equation}
\noindent
where $m_b$ is the baryonic rest mass, $k_F$=$(1.5\pi^2\rho)^{\frac{1}{3}}$ which equals Fermi momentum in case of SNM, the kinetic energy per baryon $\epsilon^{kin}$=$\left[\frac{3\hbar^2k_F^2}{10m_b}\right] F(X)$ with $F(X)$=$\left[\frac{(1+X)^{5/3} + (1-X)^{5/3}}{2}\right]$ and $J_v$=$J_{v00} + X^2 J_{v01}$, $J_{v00}$ and $J_{v01}$ represent the volume integrals of the isoscalar and the isovector parts of the M3Y interaction. The isoscalar $t_{00}^{M3Y}$ and the isovector $t_{01}^{M3Y}$ components of M3Y interaction potential are given by

\begin{eqnarray}
 t_{00}^{M3Y}(s, \epsilon^{kin})=&&+7999\frac{\exp( - 4s)}{4s}-2134\frac{\exp( -2.5s)}{2.5s} \nonumber \\
 && +J_{00}(1-\alpha\epsilon^{kin}) \delta(s) \nonumber \\ 
 t_{01}^{M3Y}(s, \epsilon^{kin})=&&-4886\frac{\exp( - 4s)}{4s}+1176\frac{\exp( -2.5s)}{2.5s} \nonumber \\
 && +J_{01}(1-\alpha\epsilon^{kin}) \delta(s)
\label{seqn2}
\end{eqnarray} 
\noindent
where $s$ represents the relative distance between two interacting baryons, $J_{00}=-276$ MeV.fm$^3$, $J_{01}=+228$ MeV.fm$^3$ and the energy dependence parameter $\alpha=0.005$ MeV$^{-1}$. The strengths of the Yukawas were extracted by fitting its matrix elements in an oscillator basis to those elements of G-matrix obtained with Reid-Elliott soft core NN interaction and the ranges were selected to ensure OPEP tails in the relevant channels as well as a short-range part which simulates the $\sigma$-exchange process \cite{Be77}. The density dependence is employed to account for the Pauli blocking effects and the higher order exchange effects \cite{Sa79}. Thus the DDM3Y effective NN interaction is given by $v_{0i}(s,\rho, \epsilon^{kin}) = t_{0i}^{M3Y}(s, \epsilon^{kin}) g(\rho)$ where the density dependence $g(\rho) = C (1 - \beta \rho^n)$ \cite{BCS08} with $C$ and $\beta$ being the constants of density dependence.
    
    The $\beta$-equilibrated nuclear matter EoS is obtained by evaluating the asymmetric nuclear matter EoS at the isospin asymmetry $X=1 - 2 x_p$ determined from the $\beta$-equilibrium proton fraction $x_p$ [$=\frac{\rho_p}{\rho}$], obtained by solving $\hbar c (3 \pi^2\rho x_p)^{1/3} = -\frac{\partial \epsilon(\rho,x_p)}{\partial x_p} = +2\frac{\partial \epsilon}{\partial X}$. The thermodynamical method requires the system to obey the intrinsic stability condition $V_{thermal} > 0$ which is given by 
    
\begin{eqnarray}
V_{thermal}=\rho^2\left[2\rho\frac{\partial \epsilon}{\partial \rho}+\rho^2\frac{\partial^2 \epsilon}{\partial \rho^2}-\rho^2\frac{\left(\frac{\partial^2 \epsilon}{\partial \rho \partial x_p}\right)^2}{\frac{\partial^2 \epsilon}{\partial x_p^2}}\right]
\label{seqn3}
\end{eqnarray}
and obviously, it goes to zero at the inner edge separating the liquid core from the solid crust since it corresponds to a phase transition from the homogeneous matter at high densities to the inhomogeneous
matter at low densities. The core-crust transition density $\rho_t$, pressure P$_t$ and proton fraction  $x_{p(t)}$ of the neutron stars are obtained \cite{At14} by setting $V_{thermal}=0$ which goes to negative with decreasing density.

\noindent
\section{ Crustal fraction of moment of inertia in neutron stars }
\label{section3}

    The crustal fraction of the moment of inertia $\frac{\Delta I}{I}$ can be expressed as a function of $M$
(gravitational mass of the star) and $R$ (radius of the star) with the only dependence on the equation of state
arising from the values of transition density $\rho_t$ and pressure P$_t$. Actually, the major dependence is on the value of P$_t$, since $\rho_t$ enters only as a correction in the following approximate formula \cite{Li99}

\begin{eqnarray}
\frac{\Delta I}{I} \approx \frac{28\pi {\rm P}_t R^3}{3Mc^2} \left(\frac{1-1.67\xi-0.6\xi^2}{\xi}\right) \nonumber\\
\times \left(1+\frac{2{\rm P}_t}{\rho_t m_b c^2}\frac{(1+7\xi)(1-2\xi)}{\xi^2} \right)^{-1}
\label{seqn4}
\end{eqnarray} 
where $\xi=\frac{GM}{Rc^2}$. The crustal fraction of the moment of inertia is particularly interesting as it can be inferred from observations of pulsar glitches, the occasional disruptions of the otherwise extremely regular pulsations from magnetized, rotating neutron stars \cite{Xu09a}. Link et al. \cite{Li99} showed that glitches represent a self-regulating instability for which the star prepares over a waiting time. The angular momentum requirements of glitches in the Vela pulsar indicate that more than 0.014 of the moment of inertia drives these events. So, if glitches originate in the liquid of the inner crust, this means that $\frac{\Delta I}{I} > 1.4\%$.

\noindent
\section{Tolman-Oppenheimer-Volkoff Equation and mass-radius relation}
\label{section4}

    In astrophysics, the Tolman-Oppenheimer-Volkoff (TOV) equation \cite{TOV39a,TOV39b} constrains the structure of a spherically symmetric body of isotropic material which is in static gravitational equilibrium, as modelled by general relativity and is given by

\vspace{-0.0cm}
\begin{eqnarray}
\frac{dP(r)}{dr} = -\frac{G}{c^4}\frac{[\varepsilon(r)+P(r)][m(r)c^2+4\pi r^3P(r)]}{r^2[1-\frac{2Gm(r)}{rc^2}]} \\ 
{\rm where} ~\varepsilon(r)=(\epsilon + m_b c^2)\rho(r),~m(r)c^2=\int_0^r \varepsilon(r') d^3r' \nonumber
\label{seqn5}
\end{eqnarray}
\noindent   
which can be easily solved numerically using Runge-Kutta method for masses and radii. The quantities $\varepsilon(r)$ and $P(r)$ are the energy density and pressure at a radial distance $r$ from the centre, and are given by the equation of state. The mass of the star contained within a distance $r$ is given by $m(r)$. The size of the star is determined by the boundary condition $P(r)=0$ and the total mass $M$ of the star integrated up to the surface $R$ is given by $M=m(R)$ \cite{Um97}. The single integration constant needed to solve the TOV equation is $P_c$, the pressure at the center of the star calculated at a given central density $\rho_c$. The masses of slowly rotating neutron stars are very close \cite{Ch10,Mi12,Ba14} to those obtained by solving TOV equation.   
    
\begin{figure}[t]
\vspace{0.0cm}
\eject\centerline{\epsfig{file=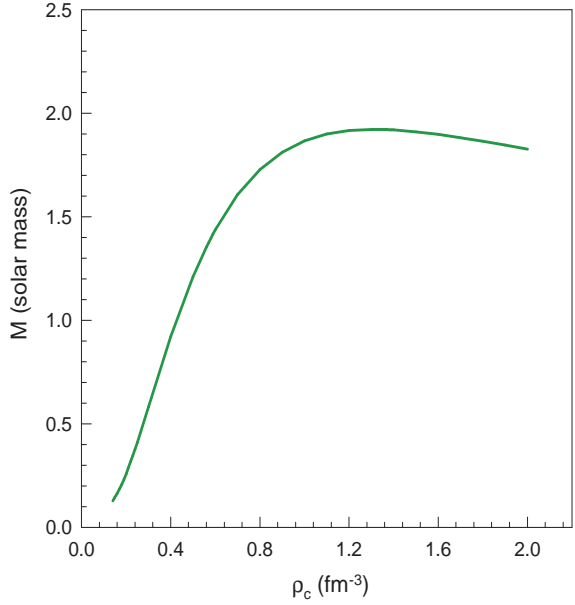,height=8cm,width=7.5cm}}
\caption
{Variation of mass with central density for slowly rotating neutron stars for the present nuclear EoS.}
\label{fig1}
\vspace{0.0cm}
\end{figure}
\noindent 
   
\begin{figure}[t]
\vspace{0.0cm}
\eject\centerline{\epsfig{file=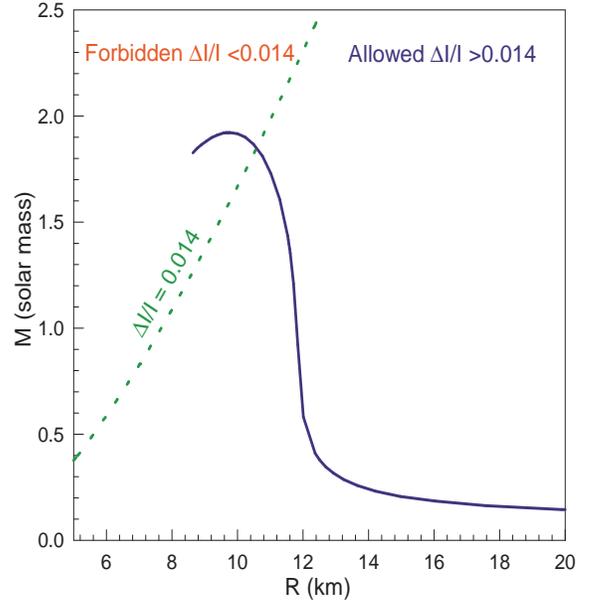,height=8cm,width=7.5cm}}
\caption
{The mass-radius relation of slowly rotating neutron stars for the present nuclear EoS. For the Vela pulsar, the constraint of $\frac{\Delta I}{I} > 1.4\%$ implies that allowed masses and radii lie to the right of the line defined by $\frac{\Delta I}{I} = 0.014$ (for $\rho_t=$ 0.0938 fm$^{-3}$, P$_t=$ 0.5006 MeV fm$^{-3}$).}
\label{fig2}
\vspace{0.0cm}
\end{figure}
\noindent 

    The moment of inertia of neutron stars is calculated by assuming the star to be rotating slowly with a uniform angular velocity $\Omega$ \cite{Ar77}. The angular velocity $\bar{\omega}(r)$ of a point in the star measured with respect to the angular velocity of the local inertial frame is determined by the equation

\vspace{-0.0cm}
\begin{equation}
\frac{1}{r^4}\frac{d}{dr}\left[r^4 j \frac{d\bar{\omega}}{dr}\right] + \frac{4}{r}\frac{dj}{dr}\bar{\omega}= 0
\label{seqn6}
\end{equation}
\noindent 
where

\vspace{-0.0cm}
\begin{equation}
j(r) = e^{-\phi(r)} \sqrt{1-\frac{2Gm(r)}{rc^2}}.
\label{seqn7}
\end{equation}
\noindent 
The function $\phi(r)$ is constrained by the condition

\vspace{-0.0cm}
\begin{equation}
e^{\phi(r)}\mu(r)={\rm constant}=\mu(R)\sqrt{1-\frac{2GM}{Rc^2}}
\label{seqn8}
\end{equation}
\noindent 
where the chemical potential $\mu(r)$ is defined as

\vspace{-0.0cm}
\begin{equation}
\mu(r) = \frac{\varepsilon(r)+P(r)}{\rho(r)}.
\label{seqn9}
\end{equation}
\noindent
Using these relations, Eq.(6) can be solved subject to the boundary conditions that $\bar{\omega}(r)$ is regular as $r \rightarrow 0$ and $\bar{\omega}(r) \rightarrow \Omega$ as $r \rightarrow \infty$. Then moment of inertia of the star can be calculated using the definition $I = J/\Omega$, where the total angular momentum $J$ is given as

\vspace{-0.0cm}
\begin{equation}
J = \frac{c^2}{6G} R^4 \frac{d\bar{\omega}}{dr}\Big|_{r= R}.
\label{seqn10}
\end{equation}
\noindent
       
\begin{table}[htbp]
\centering
\caption{Results of present calculations for $n$=$\frac{2}{3}$ of symmetric nuclear matter incompressibility $K_\infty$, nuclear symmetry energy at saturation density $E_{sym}(\rho_0)$, the slope $L$ and  isospin dependent part $K_\tau$ of the isobaric incompressibility (all in MeV) \cite{Ba09} are tabulated along with the density, pressure and proton fraction at the core-crust transition for $\beta$-equilibrated neutron star matter and corresponding Vela pulsar constraint.}
\begin{tabular}{cccc}
\hline
\hline
$K_\infty$&$E_{sym}(\rho_0)$&$L$&$K_\tau$ \\ 
\hline
 $274.7\pm7.4$&$30.71\pm0.26$&$45.11\pm0.02$&$-408.97\pm3.01$ \\
\hline
$\rho_t$(fm$^{-3}$)& P$_t$(MeVfm$^{-3}$) & $x_{p(t)}$&Vela pulsar R(km) \\
\hline
0.0938&0.5006& 0.0308&$R \geq 4.10 + 3.36 M/M_\odot$ \\
\hline
\hline
\end{tabular} 
\label{table1}
\end{table}
\noindent 
   
\begin{table}[htbp]
\centering
\caption{Radii, masses, total $\&$ crustal fraction of moment of inertia and crustal thickness as functions of central density $\rho_c$.}
\begin{tabular}{||c|c|c|c|c|c||}
\hline 
\hline
~~~~$\rho_c$~~~~&~~~~$R$~~~~&~~~~$M$~~~~&~~~~$I$~~~~&~~~~$\frac{\Delta I}{I}$~~~~&~~~~$\Delta R$~~~~\\ \hline

fm$^{-3}$ & km & $M_\odot$ & $M_\odot$km$^2$ & fraction & km\\ \hline
\hline
 2.00&   8.6349&   1.8277&    70.88&    0.0055&          0.2462 \\
 1.90&   8.7598&   1.8467&    73.83&    0.0057&          0.2523 \\
 1.80&   8.8957&   1.8651&    77.00&    0.0060&          0.2598 \\
 1.70&   9.0444&   1.8824&    80.38&    0.0063&          0.2696 \\
 1.60&   9.2052&   1.8980&    83.97&    0.0067&          0.2806 \\
 1.50&   9.3810&   1.9109&    87.70&    0.0072&          0.2951 \\
 1.40&   9.5710&   1.9197&    91.52&    0.0079&          0.3121 \\
 1.39&   9.5911&   1.9203&    91.91&    0.0080&          0.3144 \\
 1.38&   9.6109&   1.9208&    92.29&    0.0080&          0.3161 \\
 1.37&   9.6314&   1.9213&    92.67&    0.0081&          0.3185 \\
 1.36&   9.6514&   1.9217&    93.05&    0.0082&          0.3203 \\
 1.35&   9.6718&   1.9220&    93.43&    0.0083&          0.3222 \\
 1.34&   9.6928&   1.9223&    93.81&    0.0084&          0.3248 \\
 1.33&   9.7141&   1.9225&    94.18&    0.0085&          0.3275 \\
 1.32&   9.7349&   1.9226&    94.55&    0.0085&          0.3296 \\
 1.31&   9.7559&   1.9227&    94.93&    0.0086&          0.3318 \\
 1.30&   9.7770&   1.9226&    95.30&    0.0087&          0.3340 \\
 1.20&   9.9995&   1.9173&    98.85&    0.0098&          0.3620 \\
 1.10&  10.2371&   1.9004&   101.88&    0.0112&          0.3970 \\
 1.00&  10.4902&   1.8675&   103.94&    0.0132&          0.4441 \\
 0.90&  10.7544&   1.8127&   104.42&    0.0158&          0.5066 \\
 0.80&  11.0239&   1.7285&   102.47&    0.0197&          0.5929 \\
 0.70&  11.2865&   1.6064&    97.04&    0.0255&          0.7148 \\
 0.60&  11.5245&   1.4369&    87.06&    0.0344&          0.8952 \\
 0.59&  11.5456&   1.4170&    85.78&    0.0356&          0.9175 \\
 0.58&  11.5666&   1.3965&    84.44&    0.0368&          0.9411 \\
 0.57&  11.5874&   1.3753&    83.04&    0.0381&          0.9663 \\
 0.56&  11.6073&   1.3536&    81.58&    0.0394&          0.9924 \\
 0.55&  11.6262&   1.3313&    80.07&    0.0408&          1.0193 \\
 0.50&  11.7135&   1.2104&    71.65&    0.0492&          1.1792 \\
 0.45&  11.7830&   1.0734&    61.88&    0.0602&          1.3897 \\
 0.40&  11.8388&   0.9206&    51.00&    0.0752&          1.6801 \\
 0.30&  12.0129&   0.5808&    28.54&    0.1249&          2.7618 \\
 0.25&  12.3703&   0.4103&    19.24&    0.1686&          3.9149 \\
 0.24&  12.5113&   0.3779&    17.73&    0.1805&          4.2542 \\
 0.23&  12.6944&   0.3464&    16.35&    0.1942&          4.6511 \\
 0.22&  12.9314&   0.3159&    15.14&    0.2103&          5.1189 \\
 0.21&  13.2434&   0.2867&    14.12&    0.2296&          5.6802 \\
 0.20&  13.6576&   0.2587&    13.31&    0.2537&          6.3643 \\
 0.19&  14.2131&   0.2323&    12.74&    0.2847&          7.2125 \\
 0.18&  14.9725&   0.2075&    12.47&    0.3265&          8.2904 \\
 0.17&  16.0398&   0.1845&    12.59&    0.3863&          9.7057 \\
 0.16&  17.5771&   0.1634&    13.25&    0.4767&         11.6254 \\
 0.15&  19.8913&   0.1445&    14.77&    0.6254&         14.3634 \\
 0.14&  23.5740&   0.1278&    17.88&    0.8972&         18.5215 \\ \hline 
\hline
\end{tabular}
\label{table3} 
\end{table}
\noindent 

\begin{table*}[htbp]
\centering
\caption{Variations of the core-crust transition density, pressure and proton fraction for $\beta$-equilibrated neutron star matter, symmetric nuclear matter incompressibility $K_\infty$ and isospin dependent part $K_\tau$ of isobaric incompressibility with parameter $n$.}
\begin{tabular}{||c|c|c|c|c|c|c|c|c|c||}
\hline 
\hline
$n$&$\rho_t$& P$_t$ & x$_{p(t)}$&$K_\infty$&$K_\tau$& Maximum mass &~~Radius~~&Crustal Thickness&Vela Pulsar Radius Constraint \\ \hline
\hline
&fm$^{-3}$&MeVfm$^{-3}$& &MeV&MeV&$M_\odot$ &km &km &km \\ \hline
Expt.&values&- - $\rightarrow$&$\rightarrow$&250-270 &-370$\pm$120 &1.97$\pm$0.04 & & & \\ \hline
1/6&0.0797 &0.4134 & 0.0288&182.13&-293.42&1.4336&8.5671&0.4009&$R \geq 4.48 + 3.37 M/M_\odot$ \\ 
1/3&0.0855 &0.4520 & 0.0296&212.98&-332.16&1.6002&8.9572&0.3743&$R \geq 4.31 + 3.36 M/M_\odot$ \\ 
1/2&0.0901 &0.4801 & 0.0303&243.84&-370.65&1.7634&9.3561&0.3515&$R \geq 4.19 + 3.36 M/M_\odot$ \\ 
2/3&0.0938 &0.5006 & 0.0308&274.69&-408.97&1.9227&9.7559&0.3318&$R \geq 4.10 + 3.36 M/M_\odot$ \\ 
1  &0.0995 &0.5264 & 0.0316&336.40&-485.28&2.2335&10.6408&0.3088&$R \geq 3.99 + 3.36 M/M_\odot$ \\ \hline
\hline
\end{tabular}
\label{table2} 
\end{table*}
\noindent  

\noindent
\section{ Theoretical Calculations }
\label{section5}
    
    The calculations are performed using the values of the saturation density $\rho_0$=0.1533 fm$^{-3}$ \cite{Sa89} and the saturation energy per baryon $\epsilon_0$=-15.26 MeV \cite{CB06} for the SNM obtained from the co-efficient of the volume term of Bethe-Weizs\"acker mass formula which is evaluated by fitting the recent experimental and estimated atomic mass excesses from Audi-Wapstra-Thibault atomic mass table \cite{Au03} by minimizing the mean square deviation incorporating correction for the electronic binding energy \cite{Lu03}. In a similar recent work, including surface symmetry energy term, Wigner term, shell correction and proton form factor correction to Coulomb energy also, $a_v$ turns out to be 15.4496 MeV and 14.8497 MeV when $A^0$ and $A^{1/3}$ terms are also included \cite{Ro06}. Using the usual value of $\alpha=0.005$ MeV$^{-1}$ for the parameter of energy dependence of the zero range potential and $n=\frac{2}{3}$, the values obtained for the constants of density dependence $C$ and $\beta$ and the SNM incompressibility $K_\infty$ are 2.2497, 1.5934 fm$^2$ and 274.7 MeV, respectively. The saturation energy per baryon is the volume energy coefficient and the value of -15.26$\pm$0.52 MeV covers, more or less, the entire range of values obtained for $a_v$ for which now the values of $C$=2.2497$\pm$0.0420, $\beta$=1.5934$\pm$0.0085 fm$^2$ and the SNM incompressibility $K_\infty$=274.7$\pm$7.4 MeV \cite{BCS08}.  
    
    The stability of the $\beta$-equilibrated dense matter in neutron stars is investigated and the location of the inner edge of their crusts and core-crust transition density and pressure are determined using the DDM3Y effective nucleon-nucleon interaction. The results for the transition density, pressure and proton fraction at the inner edge separating the liquid core from the solid crust of neutron stars are calculated and presented in Table-I for $n=\frac{2}{3}$. The symmetric nuclear matter incompressibility $K_\infty$, nuclear symmetry energy at saturation density $E_{sym}(\rho_0)$, the slope $L$ and  isospin dependent part $K_\tau$ of the isobaric incompressibility are also tabulated since these are all in excellent agreement with the constraints recently extracted from measured isotopic dependence of the giant monopole resonances in even-A Sn isotopes, from the neutron skin thickness of nuclei, and from analyses of experimental data on isospin diffusion and isotopic scaling in intermediate energy heavy-ion collisions. 
    
    The calculations for masses and radii are performed using the EoS covering the crustal region of a compact star which are Feynman-Metropolis-Teller (FMT) \cite{FMT49}, Baym-Pethick-Sutherland (BPS) \cite{BPS71} and Baym-Bethe-Pethick (BBP) \cite{BBP71} upto number density of 0.0582 fm$^{-3}$ and $\beta$-equilibrated neutron star matter beyond. The values of $I$ obtained by solving Eq.(6) subject to the boundary conditions stated earlier are listed in Table-II along with masses $M$, radii $R$ and crustal thickness $\Delta R$ of neutron stars. Once masses and radii are determined, $\frac{\Delta I}{I}$ are obtained from Eq.(4) and listed in Table-II. In Fig.-1, variation of mass with central density is plotted for slowly rotating neutron stars for the present nuclear EoS. In Fig.-2, the mass-radius relation of slowly rotating neutron stars is shown. Using Eq.(4) again the mass-radius relation is obtained for fixed values of $\frac{\Delta I}{I}$, $\rho_t$ and P$_t$. This is then plotted in the same figure for $\frac{\Delta I}{I}$ equal to 0.014. For Vela pulsar, the constraint $\frac{\Delta I}{I}>1.4\%$ implies that allowed mass-radius lie to the right of the line defined by $\frac{\Delta I}{I} = 0.014$ (for $\rho_t$ = 0.0938 fm$^{-3}$, P$_t$ = 0.5006 MeV fm$^{-3}$). This condition is given by the inequality $R \geq 4.10 + 3.36 M/M_\odot$ kms.

    In Table-III, the variations with parameter $n$ of the core-crust transition density, pressure and proton fraction for $\beta$-equilibrated neutron star matter, symmetric nuclear matter incompressibility $K_\infty$, isospin dependent part $K_\tau$ of isobaric incompressibility, neutron star's maximum mass with corresponding radius and crustal thickness are listed along with corresponding Vela pulsar constraints. It is important to mention here that recent observations of the binary millisecond pulsar J1614-2230 by P. B. Demorest et al. \cite{De10} suggest that the masses lie within 1.97$\pm$0.04 M$_\odot$ where M$_\odot$ is the solar mass. 

\noindent
\section{ Results and discussion }
\label{section6}

    Recently, it is conjectured that the glitches observed in the Vela pulsar require an additional reservoir of angular momentum and the crust may not be enough to explain the phenomenon \cite{An12}. Large pulsar frequency glitches can be interpreted as sudden transfers of angular momentum between the neutron superfluid permeating the inner crust and the rest of the star. In spite of the absence of viscous drag, the neutron superfluid is strongly coupled to the crust due to non-dissipative entrainment effects. It is often argued that these effects may limit the maximum amount of angular momentum that can possibly be transferred during glitches \cite{Ch13}. We find that the present EoS can accommodate large crustal moments of inertia and that large enough transition pressures can be generated to explain the large Vela glitches without invoking an additional angular-momentum reservoir beyond that confined to the solid crust. Our results suggest that the crust may be enough \cite{Pi14} which can be substantiated from Table-II that $\frac{\Delta I}{I} > 0.014$ for pulsars with masses 1.8 M$_\odot$ or less. 

    The results listed in Table-III suggest that SNM incompressibility do have some effect in determining the crustal fraction of moment of inertia and on the Vela Pulsar Radius Constraint like some other recent studies \cite{St15}. But the incompressibility values of about 15 MeV window around 270 MeV corresponding to $n=\frac{2}{3}$ is experimentally supported. The present status of experimental determination of the SNM incompressibility from the compression modes ISGMR and isoscalar giant dipole resonance (ISGDR) of nuclei infers \cite{Sh06} that due to violations of self consistency in HF-RPA calculations of the strength functions of giant resonances result in shifts in the calculated values of the centroid energies which may be larger in magnitude than the current experimental uncertainties. In fact, the prediction of $K_\infty$ lying in the range of 210-220 MeV were due to the use of a not fully self-consistent Skyrme calculations \cite{Sh06}. Correcting for this drawback, Skyrme parmetrizations of SLy4 type predict $K_\infty$ values in the range of 230-240 MeV \cite{Sh06}. Moreover, it is possible to build bona fide Skyrme forces so that the SNM incompressibility is close to the relativistic value, namely 250-270 MeV. Hence, from the ISGMR experimental data, conclusion may be drawn that $K_\infty \approx$ 240 $\pm$ 20 MeV. Moreover, the constant of density dependence $\beta$=1.5934$\pm$0.0085 fm$^2$, which has the dimension of cross section for $n=\frac{2}{3}$, can be interpreted as the isospin averaged effective nucleon-nucleon interaction cross section in ground state symmetric nuclear medium. For a nucleon in ground state nuclear matter $k_F\approx$ 1.3 fm$^{-1}$ and $q_0 \sim \hbar k_F c \approx$ 260 MeV and the present result for the `in medium' effective cross section is reasonably close to the value obtained from a rigorous Dirac-Brueckner-Hartree-Fock \cite{Sa06} calculations corresponding to such $k_F$ and $q_0$ values which is $\approx$ 12 mb. Using the value of the constant of density dependence $\beta$=1.5934$\pm$0.0085 fm$^2$ corresponding to the value of the parameter $n=\frac{2}{3}$ along with the baryonic density of 0.1533 fm$^{-3}$, the value obtained for the nuclear mean free path $\lambda$ is about 4 fm which is in excellent agreement \cite{Si83} with that obtained using another method. 

\noindent
\section{ Summary and conclusion }
\label{section7}

    In summary, the DDM3Y effective interaction which provides a unified description of elastic and inelastic scattering, proton-, $\alpha$-, cluster- radioactivities and nuclear matter properties, also provides an excellent description of the $\beta$-equilibrated neutron star matter which is stiff enough at high densities to reconcile with the recent observations of the massive compact stars \cite{Ch10,Mi12,Ba14} while the corresponding symmetry energy is supersoft as preferred by the FOPI/GSI experimental data. The neutron star core-crust transition density, pressure and proton fraction determined from the thermodynamic stability condition to be $\rho_t=$ 0.0938 fm$^{-3}$, P$_t=$ 0.5006 MeV fm$^{-3}$ and $x_{p(t)}=$ 0.0308, respectively, along with observed minimum crustal fraction of the total moment of inertia of the Vela pulsar provide a new limit for its radius. It is somewhat different from the other estimates \cite{Li99,Xu09} and imposes a new constraint $R \geq 4.10 + 3.36 M/M_\odot$ kms for the mass-radius relation of Vela pulsar like neutron stars.


\begin{thebibliography}{99}

\bibitem{Ly96} A. G. Lyne in {\it Pulsars: Problems and Progress}, S. Johnston, M. A. Walker and M. Bailes, eds., 73 (ASP, 1996).

\bibitem{Li99} B. Link, R. I. Epstein and J. M. Lattimer, Phys. Rev. Lett. {\bf 83}, 3362 (1999).
 
\bibitem{Ep92} R. I. Epstein and G. Baym, Astrophys. J. {\bf 387}, 276 (1992).

\bibitem{Al93} M. A. Alpar, H. F. Chau, K. S. Cheng and D. Pines, Astrophys. J. {\bf 409}, 345 (1993).

\bibitem{Li96}  B. Link and R. I. Epstein, Astrophys. J. {\bf 457}, 844 (1996).

\bibitem{Ru98} M. Ruderman, T. Zhu, and K. Chen, Astrophys. J. {\bf 492}, 267 (1998).

\bibitem{Se99} A. Sedrakian and J. M. Cordes, Mon. Not. R. Astron. Soc. {\bf 307}, 365 (1999).

\bibitem{Ch09} P. Roy Chowdhury, D. N. Basu and C. Samanta, Phys. Rev. {\bf C 80}, 011305(R) (2009).

\bibitem{Ba09} D. N. Basu, P. Roy Chowdhury and C. Samanta, Phys. Rev. {\bf C 80}, 057304 (2009).

\bibitem{At14} Debasis Atta and D. N. Basu, Phys. Rev. {\bf C 90}, 035802 (2014). 

\bibitem{La07} J. M. Lattimer and M. Prakash, Phys. Rep. {\bf 442}, 109 (2007).

\bibitem{Ku04} S. Kubis, Phys. Rev. {\bf C 70}, 065804 (2004). 

\bibitem{Ku07} S. Kubis, Phys. Rev. {\bf C 76}, 025801 (2007).

\bibitem{Wo08} A. Worley, P. G. Krastev and B. A. Li, Astrophys. J. {\bf 685}, 390 (2008).

\bibitem{Ca85} H. B. Callen, {\it Thermodynamics and An Introduction to Thermostatistics}, 2nd edition, John Wiley $\&$ Sons, New York (1985).

\bibitem{TOV39a} R. C. Tolman, Phys. Rev. {\bf 55}, 364 (1939).

\bibitem{TOV39b} J. R. Oppenheimer and G. M. Volkoff Phys. Rev. {\bf 55}, 374 (1939).

\bibitem{La62} A. M. Lane, Nucl. Phys. {\bf 35}, 676 (1962). 

\bibitem{Sa83} G. R. Satchler, {\it Int. series of monographs on Physics}, Oxford University Press, {\it Direct Nuclear reactions}, 470 (1983).

\bibitem{BCS08} D. N. Basu, P. Roy Chowdhury and C. Samanta, Nucl. Phys. {\bf A 811}, 140 (2008).

\bibitem{Be77} G. Bertsch, J. Borysowicz, H. McManus, W. G. Love, Nucl. Phys. {\bf A 284}, 399 (1977).

\bibitem{Sa79} G. R. Satchler and W. G. Love, Phys. Reports {\bf 55}, 183 (1979). 

\bibitem{Xu09a} J. Xu, L. W. Chen, B. A. Li and H. R. Ma, Astrophys. J. {\bf 697}, 1549 (2009).

\bibitem{Um97} V. S. Uma Maheswari, D. N. Basu, J. N. De and S. K. Samaddar, Nucl. Phys. {\bf A 615}, 516 (1997).

\bibitem{Ch10} P. R. Chowdhury, A. Bhattacharyya, D. N. Basu, Phys. Rev. {\bf C 81}, 062801(R) (2010).

\bibitem{Mi12} Abhishek Mishra, P. R. Chowdhury and  D. N. Basu, Astropart. Phys. {\bf 36}, 42 (2012).

\bibitem{Ba14} D. N. Basu, Partha Roy Chowdhury and Abhishek Mishra, Eur. Phys. J. Plus {\bf 129}, 62 (2014). 

\bibitem{Ar77} W. D. Arnett and R. L. Bowers, Astrophys. J. Suppl. {\bf 33}, 415 (1977).

\bibitem{Sa89} C. Samanta, D. Bandyopadhyay and J. N. De, Phys. Lett. {\bf B 217}, 381 (1989). 

\bibitem{CB06} P. Roy Chowdhury and D. N. Basu, Acta Phys. Pol. {\bf B 37},1833 (2006).

\bibitem{Au03} G. Audi, A. H. Wapstra and C. Thibault, Nucl. Phys. {\bf A 729}, 337 (2003).

\bibitem {Lu03} D. Lunney, J. M. Pearson and C. Thibault, Rev. Mod. Phys. {\bf 75}, 1021 (2003).

\bibitem{Ro06} G. Royer and C. Gautier, Phys. Rev. {\bf C 73}, 067302 (2006).

\bibitem{FMT49} R. P. Feynman, N. Metropolis and E. Teller, Phys. Rev. {\bf 75} (1949) 1561.

\bibitem{BPS71} G. Baym, C. J. Pethick and P. Sutherland, Astrophys. J. {\bf 170} (1971) 299.

\bibitem{BBP71} G. Baym, H. A. Bethe and C. J. Pethick, Nucl. Phys. {\bf A 175} (1971) 225.

\bibitem{De10} P. B. Demorest, T. Pennucci, S. M. Ransom, M. S. E. Roberts and J. W. T. Hessels, Nature {\bf 467}, 1081 (2010).

\bibitem{An12} N. Andersson, K. Glampedakis, W. C. G. Ho and C. M. Espinoza, Phys. Rev. Lett. {\bf 109}, 241103 (2012).

\bibitem{Ch13} N. Chamel, Phys. Rev. Lett. {\bf 110}, 011101 (2013).

\bibitem{Pi14} J. Piekarewicz, F. J. Fattoyev and C. J. Horowitz, Phys. Rev. {\bf C 90}, 015803 (2014).

\bibitem{St15} A. W. Steiner, S. Gandolfi, F. J. Fattoyev and W. G. Newton, Phys. Rev. {\bf C 91}, 015804 (2015).

\bibitem{Sh06} S. Shlomo, V. M. Kolomietz and G. Col\'o, Eur. Phys. J. {\bf A 30}, 23 (2006).

\bibitem{Sa06} F. Sammarruca and P. Krastev, Phys. Rev. {\bf C 73}, 014001 (2006).

\bibitem{Si83} B. Sinha, Phys. Rev. Lett. {\bf 50}, 91 (1983).

\bibitem{Xu09} Jun Xu, Lie-Wen Chen, Bao-An Li and Hong-Ru Ma, Phys. Rev. {\bf C 79}, 035802 (2009).

\end{thebibliography}
\end{document}